\documentstyle[preprint,eqsecnum,aps]{revtex}

\begin{document}
\newcommand{\be}{\begin{equation}}
\newcommand{\ee}{\end{equation}}
\newcommand{\bea}{\begin{eqnarray}}
\newcommand{\eea}{\end{eqnarray}}
\draft
\preprint{ }
\title{Selected Topics in the Theory of 1D Quantum Wires
}
\author{A. O. Gogolin$^{(a,b)}$
}
\address{
$^{(a)}$Institut Laue--Langevin, B.P.156 38042 Grenoble, Cedex 9, France\\
E-mail: gogolin@gaston.ill.fr\\
$^{(b)}$Landau Institute for Theoretical Physics, Moscow, 117940 Russia}
\maketitle
\begin{abstract}
A qualitative discussion of recent theoretical results in the physics
of 1D quantum wires is provided here. The consideration is mainly
focused on observable quantities, such as conductance, persistent current, and
X-Ray response functions, which are discussed in simple terms.
\vspace{9cm}
\pacs{
Lecture given at the Landau Institute Seminar, Col de Port, March 1994.}
\end{abstract}
\narrowtext

\section{Introduction}
\label{sec:introduction}

Metallic wires, used as electric current conductors, are not rare in our
daily life. These wires, however, are not those which can be called
{\em quantum wires} and therefore not those which will be discussed below,
since they are very
thick. Thus, the first thing to be explained is what does it
mean to be thick (vs narrow) from the physical point of view.
Imagine a cylindrical
wire of the length $L$ and the width $l$. Electrons are moving
inside the cylinder confined by its walls (the potential energy profile induced
by walls of the wire is also called {\em confining potential}). The
single--electron states are plane waves in the direction ($x$)
along the axis of the cylinder and are quantized in the transverse direction
($\vec{\rho}$).
The single--particle spectrum hence consists of one--dimensional (1D) bands
separated because of the quantization of the transverse motion (Fig.1). The
position of the Fermi energy, $E_{F}$, tells us how many of those bands are
filled by electrons. As regards to conventional wires, e.g. ones utilized
for making telephone lines, the number $M$ of the filled bands is huge -
typically $M \sim 10^8$. In the opposite limit, when $M \sim 1$ (the
experimental set--ups with $M=2$ and $M=4$ will be discussed below), one
speaks of 1D quantum wires. The systems, for which $M$ is $\gg 1$ but still
not macroscopic (e.g. $M \sim 10^4$), are referred to as {\em mesoscopic}
systems. We will see that the physics of mesoscopic systems is essentially
different from the physics of 1D quantum wires.

In the case $M \sim 1$ we are dealing with a 1D electron gas,
which has properties that are mainly determined by electron--electron
interactions (correlations). These correlations are much more important for
1D electron systems than for electron systems of higher dimensions.
The detailed discussion of various properties of the
1D electron gas can be found in reviews\cite{Solyom&Emery}; here I just
list the most important features. First, single--electron
excitations are not well defined and can not be treated as "quasi--particles";
their density of states is vanishing at the Fermi energy. The "true" coherent
excitations, i.e. eigenmodes of the system,  are charge and spin density
fluctuations - charge and spin sounds, which are by nature bosonic,
dynamically independent and, in general,
propagate with different velocities (so-called spin-charge separation).
Conventional single-particle excitations are complicated
combinations of these charge and spin sounds. The above properties
indicate what is called in the literature as {\em Luttinger liquid} behavior.
In fact, there are only two known "universality classes" of the
low--energy behavior of electron systems: the conventional Fermi liquid (or
the Landau--Fermi liquid) and the Luttinger liquid (or the
Tomonaga--Luttinger liquid). So, experimental studies of
the 1D electron gas are of fundamental importance.

This lecture is intended as a review rather than as an original
contribution, although some of the results presented here (mainly
concerning the persistent current phenomenon) have not been published
in regular journals. In the Introduction (Sec.I), some recent
experimental results are mentioned, the main parameters of quantum wires are
discussed, and the description of a model for the  1D electron gas in the
confining potential is given. Sec.II concerns the ground--state properties
of the interacting 1D electron gas with an impurity; the perfect
reflection (transmission) phenomenon is explained. In
Sec.III, some observable quantities
(conductance, persistent current, X-Ray response functions, etc.) are
discussed in a simple manner by making use of the perfect
reflection (transmission) concept. The Discussion (Sec.IV) is devoted to
remarks on the problem of a dynamic impurity potential,
electron spin effects, and the case of many--band wires.

\subsection{Remarks on the Experimental Situation}

Before submicron--size fabrication technology provided experimentalists
with quantum wires, experiments related to the 1D electron gas
dealt exclusively with so--called
quasi--1D materials\cite{Solyom&Emery}. Recent high resolution
photoemission experiments\cite{PE} show
evidence of a strong Fermi level broadening characteristic
for the Luttinger liquid. However, these experiments (as well as all the other
imaginable ones) are limited to relatively high temperatures ($T \sim 100K$),
since at low temperatures, when the electron hopping between different
chains becomes important, all known quasi--1D materials undergo phase
transitions. These low--temperature phase transitions (typically with the
breaking of the translation invariance and formation of a charge or spin
density wave) are of great interest and reflect, of course,
the effects of electron--electron correlations within single chains but
they should be viewed as indirect consequences of these correlations and
thus mask the "pure" Luttinger liquid behavior.

On the contrary, in quantum wires,
the Luttinger liquid properties have a chance to
be directly measured. Recent experiments along this line
\cite{CSM} were done on $GaAs/AlGaAs$ quantum wires of the width
$\sim 700$ {\AA}
(in turn arranged in a planar superstructure of the period $2000$ {\AA}
within the $250$-\AA -thick $GaAs$ quantum well). The charge and spin sound
modes propagating with different velocities were detected by
resonant inelastic light scattering (the measured energy spacing between
lowest 1D
bands is  $\simeq 5meV$ and the Fermi energy was estimated to be
$E_F \simeq 6meV$, so this experimental
set--up corresponds to $M=2$ or, possibly, to $M=1$).
These experiments thus confirm theoretical predictions concerning
coherent excitations in the 1D electron gas.

Great efforts were taken to measure the conductance of quantum wires
\cite{Cond,art,NL}. Here the interest was mainly focused on the phenomenon
called {\em periodic conductance oscillations} (the devices exhibiting
those oscillations are referred to as
"single--electron transistors"\cite{Kastner}).
Although periodic conductance oscillations are somewhat outside the
material of this lecture, this provides a good point to discuss the effects of
impurities (discorder) in quantum wires. The main parameters of the 1D wires
in the experiments \cite{Cond,art,NL} were in the
order of magnitude the same as
in the experiments \cite{CSM} described above;
additionally the presence of the metallic gate
allowed one to change the electron concentration in wires by varying the gate
voltage. As described in the review paper \cite{Kastner}, it was
experimentally established that the periodic conductance oscillations
are due to the change in the number of electrons between the {\em two}
impurities which are the {\em only} ones responsible for the
effective electron scattering along the wire (in a given device).
[The experiments \cite{art},
in which these two impurities were artificially built into the devices,
provide us with a very convincing proof of this explanation.]

In general, the geometry of 1D quantum wires and known properties of
materials used for their fabrication (mainly $Si$ and $GaAs/AlGaAs$)
imply that the number of impurities effective along a wire is typically
of the order of unity. Equivalently, the electron mean free path $l_m$ is
of the order of the length of the wire. This situation should be contrasted
to what is characteristic for conventional mesoscopic systems, where
one normally has $l_m \ll L$ (and $l_m \sim l$), so that the electron motion
is dominated by the impurity scattering and hence is diffusive;
the coherent effects are thus exponentially suppressed in the
parameter $l_m /L$.
On the other hand, in quantum wires, the electron states are
(strongly) correlated
along the wire (and between each other). The relevant theoretical
question in the physics of quantum wires therefore is:
what are the properties of the interacting 1D electron gas with a single
impurity (or with several but not dense impurities)?

Regarding to the conductance, we will see that it is predicted to vanish
with temperature under above conditions. To the best of my knowledge,
this prediction is not yet convincingly confirmed by experiments; although
a related phenomenon, i.e. an essentially nonlinear current--voltage
characteristics, has been observed\cite{NL}.

Another group of interesting experiments is formed by the measurements
of the persistent current (resulting from the Aharonov--Bohm
effect for many--body systems).
Although the persistent current was predicted several decades ago\cite{early},
in the normal state systems it was discovered
by the measurements of the magnetic response of many ($\sim 10^7$) mesoscopic
rings only in 1990 \cite{Levy}.
Later, the measurements were also performed on isolated
rings\cite{single}. Both the experiments \cite{Levy} and \cite{single}
were realized on thick mesoscopic rings
with a large number of populated 1D bands
($M \sim 10^4$) and strong disorder ($l_m \sim 10^{-3}L$).
Recently measurements of the persistent current in quantum wires
($GaAs/AlGaAs$ based, with $M\simeq 4$ and a {\em weak} disorder:
$l_m \simeq 1.3L$) have been carried out\cite{Benoit}.

\subsection{Model Description of the System}

The wave--function $\Phi ( x, \vec{\rho})$ of a single electron in a wire
satisfies the Schr\"{o}dinger equation:
\begin{equation}
\left\{ -\frac{\hbar^2}{2m^*} \left( \frac{\partial^2}{\partial x^2}
+\bigtriangleup _{\displaystyle \vec{\rho}} \right)
+U_{conf}(\vec{\rho} ) \right\} \Phi \left[ x, \vec{\rho} \right]
=E\Phi ( x, \vec{\rho} )  \; ,
\label{schrodinger}
\end{equation}
where $m^*$ is the effective electron mass
(in what follows  $\hbar =1$ is assumed)
and $U_{conf}(\vec{\rho} )$ is
the confining potential. The eigenvalues and eigenfunctions of the
Eq.(\ref{schrodinger}) are:
\begin{equation}
E=\varepsilon_n +\frac{p^2}{2m^*}\; ; \; \;\; \;
\Phi ( x, \vec{\rho})= \varphi_n (\vec{\rho})\psi_{p} (x),\;\;\;
\psi_{p} (x)=e^{ipx}\; .
\label{eigen}
\end{equation}
Here $p$ is the quasimomentum in the longitudinal direction and the functions
$\varphi_n (\vec{\rho})$ correspond to the quantization of the electron
motion in the transverse direction; index $n$ labels the 1D electron bands
($n=0$ is chosen to label the lowest 1D band).

We assume now the 1D limit, i.e. only the lowest band is filled by
electrons (this means that $\varepsilon_0 < E_F < \varepsilon_1$). Then the
hamiltonian of the electron gas can be written as:
\begin{equation}
H_{bulk}=\int dx \psi^{\dagger}(x)
\left\{ -\frac{1}{2m^*} \frac{\partial^2}{\partial x^2} \right\}
\psi^{\phantom{\dagger}}(x) +
\frac{1}{2}\int dxdy
\psi^{\dagger}(x)\psi^{\dagger}(y)U(x-y)
\psi^{\phantom{\dagger}}(y)\psi^{\phantom{\dagger}}(x) \; ,
\label{ham}
\end{equation}
where $\psi^{\dagger}(x)$ and $\psi^{\phantom{\dagger}}(x)$ are electron
creation and annihilation operators referred to the states $\delta (x)
\varphi_0 (\vec{\rho})$, and the electron--electron interaction potential
$U(x)$ is the matrix element of the (three-dimensional) screened Coulomb
interaction $U(x,\vec{\rho})$:
\begin{equation}
U(x)=\int d\vec{\rho_1} d\vec{\rho_2} U(x,\vec{\rho_1}-\vec{\rho_2})
|\varphi_0 (\vec{\rho_1})|^2 |\varphi_0 (\vec{\rho_2})|^2 \; .
\label{matrixel}
\end{equation}
There are, of course, other matrix elements of the three--dimensional
interaction involving higher bands. These matrix elements are omitted,
since they are irrelevant for the low--energy properties that we are
interested in (their role is just to renormalize effective parameters).
We will also not need to specify the functions $\varphi_n (\vec{\rho})$.
Numerical calculations in this direction may be found in \cite{transfer}.

The Hamiltonian $H_{bulk}$ describes the translation invariant interacting
electron gas (that is why the subscript "bulk" is used). The impurity
scattering, also restricted to the lowest band, takes the form:
\begin{equation}
H_{im}=\int dx V(x) \psi^{\dagger}(x)\psi^{\phantom{\dagger}}(x).
\label{imp}
\end{equation}
As discussed above, we will be interested in the case when the
potential $V(x)$ corresponds to a single scatterer. For the point single
scatterer at the origin one would have $V(x)=V_0 \delta (x)$. The case with
several scatterers is not qualitatively different provided that these
scatterers are not dense. In a more rigorous way, one can formulate the
following condition for the potential: its scattering data (i.e.
scattering phases, transmission coefficient) should be smooth functions
of the momentum on the scale $\Delta p \sim 1/L$. This is equivalent to the
condition $l_m \sim L$.

For simplicity,
I will mainly consider the case of spinless electrons.
This is, of course, a drastic simplification since real electrons have spin
(note that such a situation might still be realized in a polarizing
magnetic field).
It turns out, however, that main qualitative features and concepts
of the realistic case are readily present in the model of spinless electrons.
Specific complications of the spinning case will be briefly discussed in
the conclusion.

\section{Impurity in the Interacting 1D Electron Gas}

In the last two years this problem has drawn considerable interest of
theorists.
The interest in not only due to the fact that the problem of an impurity
in Luttinger liquid is of some practical importance for quantum wires
but it is also caused by the close relation of this problem to the problem
of a quantum dissipative particle and boundary conformal field theory.
Since the pioneering work by Kane and Fisher\cite{K&F,Mattis}, intensive
studies
were devoted to the conductance behavior\cite{K&F,K&Fbis,F&N,GL,our,AFFL},
the Wigner crystal pinning\cite{Sck}, persistent current\cite{ourbis,bigcom},
Fermi--edge singularities\cite{mykol,paras} and the Kondo effect\cite{Kondo}.

The main issue is that the electron--electron interaction renormalizes
the impurity potential either to zero ({\em perfect transmission}
of the electrons) or to infinity ({\em perfect reflection}).
In order to show, not rigorously,
but rather in a simple way, the intriguing idea of the perfect reflection,
I will first follow the method devised in papers \cite{GL} and
necessary clarifying remarks will be made in the second part of this section.

\subsection{Transmission Coefficient}

Let us start from the simplest case of non--interacting electrons scattered by
the impurity potential. The single-electron wave-functions obey simple
Schr\"{o}dinger equation
\be
\left\{ -\frac{1}{2m^*}  \frac{\partial^2}{\partial x^2} + V(x) \right\}
\psi [x] = \varepsilon \psi (x)\;\; ,
\label{schrbis}
\ee
which scattering solution can be written as:
\bea
\psi_{p}(x \to -\infty ) & = & e^{ipx}+r_0 e^{-ipx},
\nonumber\\[-0.3truecm]
&~&~~~ \label{ass}\\[-0.3truecm]
\psi_{p}(x \to +\infty ) & = & t_0 e^{ipx}.
\nonumber
\eea
Here $r_0$ and $t_0$ are the reflection and transmission amplitudes
respectively
(in general, they depend smoothly on $p$).

Since the impurity potential violates the translation invariance, the mean
electron density $\rho (x)$ is not a constant but acquires an $x$-dependence
(the well known {\em Friedel oscillation}). At large $|x|$ one can write:
\be
\delta\rho (x)\simeq\frac{1}{2\pi |x|}Im\left\{ r_0 e^{2i p_F x}\right\},
\label{friedel}
\ee
where $p_F$ is the Fermi momentum. It is important that, in the 1D electron
gas,
the Friedel oscillation is long--range: $\sim 1/|x|$ (for, e.g., 3D
electrons it decays as $\sim 1/|x|^3$ only).

The next step is to take into account the electron--electron interaction.
This interaction, at the first order, can be treated within the Hartree--Fock
approximation. The ground--state wave--function is still the Slater
determinant but composed now from single--particle wave--functions satisfying
the Schr\"{o}dinger equation with an extra potential generated by
the electron--electron interaction:
\be
\left\{ -\frac{1}{2m^*}  \frac{\partial^2}{\partial x^2} + V(x) \right\}\psi
[x]
+\int dy \delta V(x,y) \psi (y)
 = \varepsilon \psi (x)\;\; .
\label{schrbisbis}
\ee
The correction $\delta V$ to the impurity potential can easily be computed
by a "mean field" decomposition
\be
\psi^{\dagger}(x)\psi^{\dagger}(y)
\psi^{\phantom{\dagger}}(y)\psi^{\phantom{\dagger}}(x) \to
\langle \psi^{\dagger}(y)\psi^{\phantom{\dagger}}(y) \rangle
\psi^{\dagger}(x)\psi^{\phantom{\dagger}}(x) -
\langle \psi^{\dagger}(y) \psi^{\phantom{\dagger}}(x) \rangle
\psi^{\dagger}(x)\psi^{\phantom{\dagger}}(y)
\label{decomposition}
\ee
of the interaction term in the hamiltonian (\ref{ham}):
\bea
&~&~~\delta V(x,y) =  \delta (x-y) V_{H}(x) - V_{ex}(x,y) =
\nonumber\\[-0.3truecm]
&~&~~~ \label{extra}\\[-0.3truecm]
&~&~~\delta (x-y) \int dy U(x-z) \delta\rho (z) -  U(x-y)
\int_{-p_{F}}^{+p_{F}}\frac{dp}{2\pi}\psi^{*}_{p}(x)\psi^{\phantom{*}}_{p}(y),
\nonumber
\eea
where the first term represents the Hartree potential and the second term is
the exchange potential.

Within the Hartree--Fock approximation the electron gas is described by
single--electron wave-functions, so we still can consider "individual
electrons" and, therefore, the question can be asked: what is the correction
$\delta D$ to the transmission coefficient $D_0 =|t_0 |^2$ due to the
interaction? The answer can straightforwardly be found: one has just to
make the first iteration in $\delta V$ of the Eq.(\ref{schrbisbis}).
That gives the
correction $\delta\psi$ to the wave--function (\ref{ass})
in the Born approximation
and hence the correction to the transmission coefficient.
Performing this simple calculation one gets \cite{GL}:
\be
\delta D =-2\alpha D_0 (1-D_0 )\ln \left(\frac{W}{|\omega |}\right)\; ,
\label{correction}
\ee
where $\omega$ is the energy of the electron scattered by the impurity
(accounted for from the Fermi level), $W$ is the electron bandwidth and the
dimensionless interaction parameter $\alpha$ is determined by the Fourier
transform of the interaction potential $U$ as:
\be
\alpha = \frac{U(0)-U(2p_F )}{2\pi v_F}\; ,
\label{alpha}
\ee
with the Fermi velocity $v_F$.

The log--divergency of the correction (\ref{correction}) to the
transmission coefficient may be interpreted as a result of the long--range
character of the Friedel oscillation (\ref{friedel}), which in turn leads to
the log-divergent Fourier transform of the Hartree potential $V_H$ (at $2p_F$).
This explains the $U(2p_F )$ contribution to the Eq.(\ref{alpha}).
The $U(0)$ contribution to $\alpha$ comes from the exchange term in
$\delta V$ and can not be so simply interpreted though it is clearly also
due to long--range perturbations induced by the impurity in the
electron gas.
Note that it is the backscattering from the impurity which is
log--divergent. Indeed, it is easy to see from the Eq.(\ref{extra}),
that scattering processes with the $2p_F$ transferred momentum only
contribute to the Eq.(\ref{alpha}). Physically this is clear: all the
long--range perturbations caused by the impurity have $2p_F$ oscillating
character. Therefore, the forward scattering is just numerically
renormalized by the interaction (see Ref.\cite{forward}) and it is the
backscattering which is the relevant process.

The expression (\ref{correction}) in only valid for $\alpha \ln (W/|\omega |)
\ll 1$, so we are unable yet to determine the scattering of electrons with
energies close to the Fermi level: $\omega \to 0$. In order to do this
one has to go to higher orders in $\alpha$, what is, strictly speaking,
impossible since the single--electron transmission coefficient is
undefined beyond the Hartree--Fock approximation. One can nevertheless
get some more insight into the problem ignoring the latter complication
and making use of the "poor man's" scaling idea\cite{Andpm}. Following
standard prescriptions of eliminating high-energy degrees of freedom,
one obtains a flow equation for the transmission coefficient\cite{GL}:
\be
\frac{dD}{d \xi}=-2\alpha D(1-D)\; ,
\label{poor}
\ee
where the scaling parameter $\xi$ may be identified with $\ln (W/|\omega |)$
(or with $\ln (W/T)$, $T$ being the temperature).
Practically, the "poor man's" scaling equation can always be obtained just by
making a differential equation from the first order log--correction,
Eq.(\ref{correction}).
The solution of the Eq.(\ref{poor}) is:
\be
D(\omega )=\frac{D_0 |\omega /W|^{2\alpha}}{R_0
+D_0 |\omega /W|^{2\alpha}}\; .
\label{trans}
\ee

\subsection{Physical picture and scaling properties}

Thus, as follows from the
Eq.(\ref{trans}), the transmission coefficient, if it was defined,
would be the function of the energy of the shape shown in Fig.2, implying
perfect transmission for $\alpha < 0$ and perfect reflection
for $\alpha > 0$. This turns indeed out to be
qualitatively correct, although the above
arguments might look obscure at the moment. Some type of a partial summation of
the log-divergent terms is provided by the Eq.(\ref{poor}),
written for the transmission coefficient, which is well
defined at the first order in $\alpha$ only. The result, Eq.(\ref{trans}),
crucially depends on the sign of the interaction constant $\alpha$,
and the reason for this dependence is completely unclear at the
present stage. In order to stress this point it is worth noticing that
if we were seriously relied on the Hartree--Fock approximation and
calculated the transmission coefficient not just by the first iteration
but from the exact scattering solution of the Eq.(\ref{schrbisbis}), we would
find a drastically different (and wrong) result:
transmission coefficient vanishes
at the Fermi energy irrespectively of the sign of $\alpha$. Why?
One can clarify this point only taking into account the
ground--state properties
of our electron system, i.e. the Luttinger liquid properties.

Although, in 1D electron systems, the long--range order is always destroyed by
quantum fluctuations\cite{Solyom&Emery}, one should ask which correlations
are enhanced (or suppressed) in the ground--state. The long--distance behavior
of the correlation functions (in the pure case)
was intensively studied\cite{Solyom&Emery,correl}. In the simplest case of
spinless electrons there are only two correlation functions
of interest; namely the density--density function and the pair--pair function,
which large $x$ asymptotics is given by:
\be
N(x)=\langle
\psi^{\dagger}(x)
\psi^{\phantom{\dagger}}(x)\psi^{\dagger}(0)
\psi^{\phantom{\dagger}}(0) \rangle
\sim \frac{\cos (2p_F x)}{x^{2K}} \; ,
\label{dens}
\ee
\be
\Delta (x)=\langle
\psi^{\dagger}(x)\psi^{\dagger}(x)
\psi^{\phantom{\dagger}}(0)
\psi^{\phantom{\dagger}}(0) \rangle
\sim \frac{1}{x^{2/K}}\; ,
\label{pair}
\ee
where the so--called {\em exponent} $K$ is a key quantity, which completely
determines the low--energy properties of the system. A naive estimation of $K$
from the hamiltonian (\ref{ham}) gives: $K=1/\sqrt{1+2\alpha}$. In general,
the exponents are intended as phenomenological quantities (which can exactly be
determined in some particular cases only\cite{correl}).
The case $K<1$ ($\alpha >0$) corresponds to the repulsive interaction: the
charge-density wave correlations are enhanced, Eq.(\ref{dens}),
but the pair (superconductivity) correlations are suppressed, Eq.(\ref{pair}).
In the case of attractive interaction, $K>1$ ($\alpha <0$), the
charge-density
correlations are weak and the pair correlations dominate.

The extreme sensitivity of
impurity effects, Eq.(\ref{trans}), to the sign of
the interaction can now easily be understood. For $K<1$, the electron system
is establishing a charge-density wave,
which, being a density modulation in real space, is pinned
by the impurity. If there was a long--range order (i.e. classical
charge-density wave), the
impurity would completely block the motion of the electron gas. Since the
long--range order is destroied by fluctuations, the transmission coefficient
vanishes at zero energy (and zero temperature) only. On the contrary,
for $K>1$, pair correlations dominate, which do not correspond to any
modulations
in real space and therefore the electron gas is insensitive to the impurity
in this case (in fact, this is true even if there is a long--range order, since
the Cooper pairing takes
place in the states adjusted to the impurity potential
\cite{anddis}).

The right quantity, subject to the renormalization group (scaling) equation, is
simply the impurity potential V \cite{backV}:
\be
\frac{dV}{d\xi}=\beta (K,V)\; ,
\label{RG}
\ee
where the right hand side is referred to as the $\beta$-function\cite{notexp}.
For a weak
interaction $K\simeq 1$ and a small potential $V<<1$ one can deduce from
Eq.(\ref{poor}) that $\beta (K\simeq 1,V<<1)\simeq\alpha V$
(since $1-D\sim V^2$).
The calculation for an arbitrary interaction strength but a small $V$
gives\cite{K&F}:
\be
\beta (K,V<<1)=(1-K)V\; .
\label{beta}
\ee
In the opposite limit of strong imputity potential, it is convenient to study
the renormalization of the tunneling $t$ between two semi--infinite
leads\cite{K&F}. That gives:
\be
\frac{dt}{d\xi}=\left( 1-\frac{1}{K}\right) t \; .
\label{betat}
\ee
In the case of free electrons, $K=1$, the transmission coefficient is a
well--behaved function of the impurity potential; in other words the
latter is not renormalized under scaling\cite{boss} (i.e. the impurity
potential
is {\em marginal}), what implies the identically vanishing $\beta$-function:
$\beta (K=1,V)=0$.

The qualitative behavior of the $\beta$-function is shown in the Fig.3 and the
corresponding scaling flow phase diagram is drawn in the Fig.4.
This phase diagram can be understood from the point of view of the scale
invariance idea. It is well known that, at the point of the second order
phase transition, the fluctuations with all the possible length scales are
present and, therefore, any system should be scale invariant at the
criticality.
For 1D electron systems the critical temperature is zero and thus just the
zero--temperature theory is scale invariant (strictly speaking,
it is the effective low--energy theory which is expected to be scale
invariant).
The above results simply mean that although the interacting Luttinger liquid
without the impurity as well as the non--interacting Luttinger liquid
(i.e. the Fermi liquid) with
the impurity are scale invariant, the interacting Luttinger liquid with
the impurity is not. Thus, in the latter case, the system
flows under scale transformations to
the invariant fixed point at which there are only two possibilities
for the impurity potential: to be either zero or infinity. This is
also called a {\em boundary critical phenomenon}. In fact, the scale
invariance can be extended to the conformal invariance, which allows one to
apply the power of the boundary conformal field theory to these
problems\cite{AFFL}.

\section{Physical Properties of Quantum Wires}

The knowledge we gained about the ground--state of the system can now
be used for the calculation of the observable quantities (response
functions); some of them are considered in what follows.

\subsection{Conductance}

The conductance of the pure non--interacting Fermi gas is well known to be
$G_0 =e^2 /2\pi$ ({\em perfect} conductance).
In the case of several filled bands the total
conductance is simply the sum of the conductances for each band:
$G=2MG_0$ (the factor 2 results from the spin degeneracy). Thus, varying
the electron concentration (by changing the gate voltage), one should
get steps in the conductance. This behavior is known as the {\em conductance
quantization}; it was observed in the case of short wires
(i.e. in the so-called {\em ballistic} electron motion
regime, see Ref.\cite{cq}).

For long wires the impurity and interaction effects modify the conductance.
However,
being considered separately, these effects lead just to numerical
renormalizations of the conductance. The electron--electron interaction in the
pure system results (see, e.g., Ref.\cite{K&Fbis}): $G=KG_0$.
For the impure case, but
without the interactions, one has the famous Landauer formula\cite{landauer}:
\be
G=D_0 G_0\; .
\label{landauer}
\ee

It is the interplay of the interaction and the impurity, which
leads to drastic change of the conductance. Looking at the Eq.(\ref{trans}),
and assuming that the Landauer formula can still be used
(what is, strictly speaking,
only correct at the first order of the interaction), one can guess that,
for the repulsive interaction ($\alpha > 0$), the conductance, $G(\omega )$,
should vanish with the power low, $G(\omega )\sim |\omega /W|^{2\alpha}$.
Whereas for the attractive interaction ($\alpha > 0$), it should be almost
perfect with a deviation $G(\omega )-G_0 \sim |\omega /W|^{-2\alpha}$.
Apart from the exact values of the exponents, this guess is correct.
An accurate calculation of the conductance directly from the Kubo
formula gives\cite{K&F,K&Fbis}:
\be
G(\omega )= K G_0 \left\{
\begin{array}{l}
1-d_1 |\omega /W|^{2-2/K}\;\;\;\;K>1\;({\rm attraction})\\
d_2 |\omega /W|^{2/K-2}\;\;\;\;\;\;\;\;\;\;K<1\;({\rm repulsion})\\ \; ,
\end{array}
\right.
\label{cond}
\ee
where $d_1$ and $d_2$ are numerical coefficients of the order of unity.
The energy variable $\omega$ should be understood as
\be
\omega\to\max\left\{\omega ,T,eV_t \right\} \; ,
\label{omega}
\ee
with $V_t$ being external voltage. In fact,
the exponents of the Eq.(\ref{cond}) can be deduced
from the
scaling equations (\ref{RG}) and (\ref{betat}) without any calculations,
since one clearly has
$G\sim t^2$ for repulsion and $G-G_0\sim V^2$ for attraction.

The assumption concerning the short--range character of the electron--electron
interaction, which is necessary to obtain Eq.(\ref{cond}) for the
conductance, implies that the Coulomb repulsion between the electrons confined
to the wire is effectively screened (by metallic environments).
This is, however, not the case for some experimental
set-ups \cite{CSM}, where the long--range
Coulomb repulsion is expected to be effective
(at the length comparable to the length $L$ of the wire).
The interaction potential, Eq.(\ref{matrixel}), behaves then as
$U(x)\sim 1/|x|$ at large $x$ and tends to a finite constant for
$x$ smaller than the width $l$ of the wire, so its low--momentum
Fourier transform given by:
\be
U(q)=(2e^2/\kappa)
\ln{(1/|q|l)}\; ,
\label{coulfourier}
\ee
$\kappa$ being the dielectric constant. Thus, the interaction constant
$\alpha$, Eq.(\ref{alpha}), diverges whereas the exponent $K$ vanishes.
One should, therefore, consider the energy dependent exponent \cite{schulz}:
$K(\omega )\sim \ln \omega$. Given the Eq.(\ref{cond}), one can guess
that, in the Coulomb case, the conductance should follow $\ln G(\omega )
\sim \ln^{3/2} \omega$. Such a behavior has been found in the context
of Wigner crystal pinning \cite{Sck} as well as in the
framework of the Luttinger liquid picture \cite{our}:
\be
G(\omega )\sim \exp[-\nu\ln^{3/2}(T_0/\omega )]\; ,
\label{coulcond}
\ee
where $T_0$ is a cutoff (typically of the order of the bandwidth) and
$\nu= \sqrt{\pi\kappa v_F/9e^2}$. Thus, in the case of the long--range
Coulomb interaction the conductance vanishes (with frequency or
temperature) faster than any power low.
In the intermediate situation, when the screening length $l_{sc}$
satisfies the inequality $a_0 \ll l_{sc} \ll L$ ($a_0$ being the
average distance between electrons in the wire), one should
expect a crossover behavior of the conductance; interpolating between the
the Eq.(\ref{coulcond}) at high temperatures
and the Eq.(\ref{cond}) at low
temperatures.

\subsection{Minimal Effort Approach}

Here I pause to discuss the observations we made so far.
As we are convinced by the scaling analysis,
the electron--electron interaction, in the presence of the impurity,
leads either to the perfect transmission or to
the perfect reflection of electrons
(at the Fermi energy). Let us consider some low--energy
property {\bf  Q} of the electron gas (like the conductance).
For non--interacting electrons this property is determined
by the scattering data {\bf S} at the Fermi energy
of the impurity potential:
{\bf  Q}$=${\bf  Q}$(${\bf S}$)$.
Suppose now that this formula,
{\bf Q}$=${\bf Q}$(${\bf S}$)$,
is still valid for the case of interacting electrons provided that
{\bf S} is renormalized by the electron--electron interaction.
Just at the Fermi energy the interaction renormalizes
{\bf S} to a universal value {\bf S}$_{U}$
corresponding either to zero potential,
{{\bf S}$_{U}=${\bf S}$_{0}$,
or to infinite one,
{{\bf S}$_{U}=${\bf S}$_{\infty}$.
Considering a small deviation $\omega$ from the Fermi energy (or
assuming a small but finite temperature) we can estimate the deviation
$\delta${\bf S} and, therefore, the deviation
$\delta${\bf Q}$=$
{\bf Q}$(${\bf S}$_{U}$
$+\delta${\bf S}$)-$
{\bf Q}$(${\bf S}$_{U})$ of our observable
quantity from its universal value. We have seen how
this prescription works for the case of the conductance when
$G$ plays the role of {\bf Q}
and $D$ plays the role of {\bf S}, and
that it indeed leads to correct results
apart from exact values of exponents.
In the rest of this section I will describe how such a simple,
"minimal effort" approach can be applied for the calculation of some other
physical quantities (and compare the results with those obtained by
more rigorous methods).

\subsection{Persistent Current}

Consider a metallic ring penetrated by a magnetic flux $\Phi$. It is convenient
to define the dimensionless flux $\varphi =2\pi\Phi /\Phi_0$, $\Phi_0$
being the elementary flux quantum. The presence of the flux changes the
boundary conditions for the electron wave--function: it acquires the
phase $e^{i\varphi}$ under the rotation of each electron around the ring.
The ground--state energy $E_0$ depends on the flux, and the
(zero--temperature) current is defined as $j(\varphi )=-\partial E_0
(\varphi )/\varphi$. It can easily be calculated for noninteraction
electrons in a pure ring (see, e.g., Ref.\cite{cheung}).
As in the case of the conductance, the electron--electron
interaction does not qualitatively modify the results, provided
that there are no impurities \cite{loss1,loss2}.

In order to discuss the interplay of the impurity and interaction effects,
we need to have a formula (for non--interacting electrons),
which explicitly relates the current $j(\varphi )$
to the scattering potential (an analog of the Landauer formula for the
conductance). Such a formula was only recently derived\cite{ourbis},
just in the connection with present problem\cite{notefor}. It reads:
\begin{eqnarray}
&&j(\varphi )=j_{0}(\varphi )+j^{par}(\varphi );
\nonumber \\
&&j_{0}(\varphi )=\frac{e v_{_F}}{\pi L}
\frac{\sqrt{D_{0}}\sin \varphi}{\sqrt{1-D_{0} \cos^{2} \varphi}}
\cos^{-1} \left( \sqrt{D_0}\cos \varphi \right) , \label{formula}\\
&&j^{par}(\varphi )=-\frac{e v_{_F}}{L}
\frac{\sqrt{D_{0}}\sin \varphi}{\sqrt{1-D_{0} \cos^{2} \varphi}},
\;\;\; (N={\rm even}); \;\;\;\;\;\;\; j^{par}(\varphi )=0,
\;\;\; (N={\rm odd})\;,\nonumber
\end{eqnarray}
where $N$ is the number of electrons in the ring.

The derivation of the formula (\ref{formula})
may be found in the paper \cite{ourbis}; a short derivation
is given in the Appendix.
For the qualitative
discussion of the interaction effects we only need the
strong scattering limit ($D_0 \to 0$) of the formula (\ref{formula}),
which was found in the paper \cite{cheung}.

Let's now switch the electron--electron interaction on.
For the attractive interaction $D_0 \to 1$ and the impurity scattering is
not effective, i.e. the persistent current is the same as
the one for the pure
ring. For the most intriguing case of the repulsive interaction one has
$D_0 \to 0$. First of all I rewrite the Eq.(\ref{formula}) for the case of
small $D_0$ :
\be
j(\varphi )\simeq (-1)^N \;\frac{e v_F}{L}\sqrt{D_0}\sin \varphi \; ,
\label{formulabis}
\ee
and recall that, due to the finite length of the ring, the minimal
energy deviation from the Fermi level is $\Delta=2\pi v_F /L$ (energy spacing).
Thus, the minimal value of the transmission coefficient is
$D_0 \sim (\Delta /W)^{2\alpha}$  which, according to the
"minimal effort" approach, should be substituted
into the Eq.(\ref{formulabis}).
Indeed, a rigorous calculation gives\cite{ourbis}:
\be
j(\varphi )\sim (-1)^N \;\frac{e v_F}{L} \left(\frac{2\pi v_F}{LW}
\right)^{1/K-1} \sin \varphi \; .
\label{curint}
\ee
According to the Eq.(\ref{curint}), the repulsive interaction leads to a
simple, $\sin \varphi$ - like, flux dependence of the current, pronounced
parity effect [the factor $(-1)^N$], and considerable suppression of the
current amplitude due to the impurity scattering renormalization\cite{noteint}.
These qualitative conclusions are in a good agreement with numerical
calculations \cite{pcnumerical}.

For a small (as compared to the bandwidth) but finite temperature $T$,
the smearing of the Fermi level causes two effects:
(i) the current amplitude is exponentially suppressed by the factor
$\exp (-2\pi^2 T/\Delta)$, see Ref.\cite{cheung}; (ii) since
the main contribution
to the current comes from the states lying in the interval of the
width $\sim T$ in the vicinity of the Fermi level, the transmission
coefficient in the Eq.(\ref{formulabis}) should be calculated not at the
energy $\omega \sim \Delta$ as in the zero--temperature case but at the
energy $\omega \sim T$, what results in the factor $\sim(T/W)^{1/K-1}$.
Therefore, in the case of a finite temperature, the persistent current
can approximately be written as:
\be
j(\varphi )\sim (-1)^N \;\frac{e v_F}{L} \left( \frac{T}{W}
\right)^{1/K-1} \exp \left( -\frac{2\pi^2 T}{\Delta} \right) \sin \varphi \; .
\label{curtemp}
\ee
Unlike the case of non--interacting electrons,
the current temperature dependence is non--monotonic;
it exhibits a maximum at $T\sim\Delta$ (see Fig.5). For the first time,
such a maximum was predicted in the
paper \cite{bigcom}, where the influence of the magnetic flux
on pinned Wigner crystal has been studied. Thus, the maximum in
the current temperature dependence is a qualitative effect, independent of
the concrete model for the 1D electron gas, and it should be, in
principle, measurable.

The above results can not be quantitatively compared with the experimental
data \cite{Benoit}, first of all because of many--band character ($M\simeq 4$)
of studied samples (see also the Discussion).
In my opinion, in order to reach a complete understanding of the
persistent current phenomenon in quantum wires,
more theoretical and
experimental (e.g., concerning the temperature dependence of the current)
investigations are required.

\subsection{Orthogonality Catastrophe and Fermi Edge Singularities}

So far we have dealt with a static impurity potential.
There are, however, many physical problems, for which the
impurity potential should be regarded as a dynamic (time dependent)
one: $V=V(x,t)$.

The simplest problem of this kind is the problem
of the X-Ray absorption (emission)\cite{xrayrev}. In this case the impurity
potential, describing the interaction between the core--hole
created in the absorption process (or annihilated in the emission one)
and the electron gas in the quantum wire is suddenly switched on
(or switched of).
Although for simple metals the X-Ray problem is well
understood due to the asymptotically exact solution by Nozi\`{e}res and
De Dominicis \cite{ND}, in one-dimensional metals the
situation turns out to be qualitatively different.
One arrives at the results similar to those for simple metals,
provided that either the correlation effects \cite{S1}
or the backscattering from external potential \cite{forward}
are neglected, but
it is the interference of both, which
leads to an unusual behavior of response functions.

The "core--hole part" of the hamiltonian can be written as
\be
H_{ch}=\varepsilon_d d^{\dagger}d^{\phantom{\dagger}}+
\int dx V(x) \psi^{\dagger}(x)\psi^{\phantom{\dagger}}(x)
d^{\dagger}d^{\phantom{\dagger}}\; ,
\label{impubis}
\ee
where $d^{\dagger}$, $d^{\phantom{\dagger}}$ are the core--hole creation and
annihilation operators and $\varepsilon_d$ is the energy of the
core--hole level. There are two response functions of interest:
the core--hole Green function (relevant for X-Ray photoemission experiments):
\be
G(t)=-i\langle 0| d^{\phantom{\dagger}}(0) d^{\dagger}(t)|0\rangle\; ,
\label{chgreen}
\ee
and the spectral function
\be
I(t)=-i\langle 0| d^{\phantom{\dagger}}(0)\psi^{\dagger}(0)
\psi^{\phantom{\dagger}}(t) d^{\dagger}(t)|0\rangle \; ,
\label{spectral}
\ee
which describes the X-Ray absorption process, since the absorption of
one X-Ray quantum creates one electron in the conduction band and one hole
in a core level. Here $|0\rangle $ is the ground--state of the system in the
absence of the core--hole. At large times ($t>> 1/W$), the functions
(\ref{chgreen},\ref{spectral}) obey power lows:
\be
G(t\to\infty )\sim \frac{1}{t^{\alpha}}\; ,
\label{chgreenbis}
\ee
\be
I(t\to\infty )\sim \frac{1}{t^{1-\beta}}\; .
\label{spectralbis}
\ee
Therefore, in the energy representation, the absorption rate behaves as
$I(\omega )\sim(\omega -\omega_{th} )^{-\beta}$
at the absorption threshould $\omega_{th}$ (Fermi edge singularity).
For the case of sphearically symmetric potential [in 1D it means
$V(-x)=V(x)$] and spinless non--interacting  electrons,
the exponents, $\alpha$ and $\beta$, are known to be\cite{ND}:
\be
\alpha = \left( \frac{\delta_s}{\pi} \right)^2 \; ,
\label{expa}
\ee
\be
\beta = \frac{\delta_s}{\pi}
-\left( \frac{\delta_s}{\pi} \right)^2 \; ,
\label{expb}
\ee
where $\delta_s$ is the s--wave scattering phase.

The fact that the core--hole Green function $G(t)$ vanishes at large $t$ is
essentially the consequence of
the Anderson orthogonality catastrophe\cite{andcatastr}, which states that,
in the thermodynamic limit,
the ground--state of the system ($|0\rangle$) in the absence of the impurity
is orthogonal to that ($|V\rangle$) in the presence of the impurity:
\be
\langle 0|V\rangle \sim \frac{1}{N^{\alpha /2}}
\label{orthog}
\ee
($N$ is the number of particles). The relation between the orthogonality
exponent, Eq.(\ref{orthog}), and the exponent for
the core--hole Green function, Eq.(\ref{chgreenbis}),
is almost evident from the definition (\ref{chgreen}) and
can indeed be exactly proved\cite{hamann}.

For our purposes it is convenient to rewrite the exponent $\alpha$ in the
following form:
\be
\alpha =
 \frac{\delta_f^2}{2\pi^2}
-\frac{1}{2\pi^2} \left[ \tan^{-1}\sqrt{(1-D_0 )/D_0} \right]^{2} \; ,
\label{expabis}
\ee
where the forward scattering phase $\delta_f$ is introduced; it is defined
by $t_0 =\sqrt{D_0}\exp (-i\delta_f )$.

To compute the exponents for interacting electrons,
in a spirit of the "minimum effort" approach,
we have to impose either $D_0\to 0$ or
$D_0\to 1$. Unlike the cases of the conductance and the persistent current,
in the case of Fermi edge singularities more knowledge about scattering
data is required; in addition to $D_0$ we have to know the forward
scattering phase. As discussed above, forward scattering is not very
sensitive to the interaction. It is just numerically renormalized
\cite{forward}:
\be
\delta_f\to K^{3/2}\delta_f\; .
\label{deltaren}
\ee
Thus, for interacting electrons, the orthogonality exponent is given by:
\be
\alpha =K^{3}\frac{\delta_f^2}{2\pi^2}+\left\{
\begin{array}{l}
0\;\;\;\;\;({\rm attraction})\\
\frac{1}{8}\;\;\;\;\;({\rm repulsion})\\
\end{array}
\right.
\label{ortexp}
\ee
and the exponent $\beta$ (evidently related to $\alpha$) can be written as:
\be
\beta =K^{3/2}\frac{\delta_f}{2\pi}-K^{3}\frac{\delta_f^2}{2\pi^2} +\left\{
\begin{array}{l}
0\;\;\;\;\;({\rm attraction})\\
\frac{3}{8}\;\;\;\;\;({\rm repulsion})\\
\end{array}
\right.
\label{emiexp}
\ee

The above results were found in papers \cite{mykol} by a type of
variational method and confirmed in the paper \cite{paras} by
scaling arguments\cite{notaexp}.
Note, that response functions obey power lows
Eqs(\ref{chgreen},\ref{spectral}) with the exponents
(\ref{ortexp},\ref{emiexp}) on the largest time scale only\cite{notetime}.
The intermediate--time behavior of response functions is a quite
interesting issue but I shall not discuss it here;
the readers are referred to Refs\cite{mykol,paras}.

Generally speaking, expressions (\ref{ortexp},\ref{emiexp}) suggest that
a repulsive electron--electron interaction leads to pronounced
Fermi edge singularities in 1D metals, enhanced as compared to those in
systems of higher dimension.
Experimental evidence for such an enhancement may be found
in the paper \cite{noterecoil},
although some questions concerning the interpretation of data presented there
are not quite clear to me (for instance the role of the
core--hole recoil). The experimental determination of
Fermi edge singularities in quantum wires seems to be a
rather delicate task.
The reason is that the Fermi energy in quantum wires is very small
(of the order of several $meV$), and, in order to measure pronounced
Fermi edge singularities, the Fermi energy should be still
larger than the natural width of the core--hole level
(which is hardly possible).

\section{Discussion}

In many cases one has to consider a dynamic degree of freedom associated
with the impurity potential.
This happens in the case of the Kondo problem, which has been studied
so far by means of the "poor man's" scaling approach \cite{Kondo}, although
the knowledge of the response functions allows one to apply
more powerful methods \cite{power} to the Kondo problem. This has not yet been
done.
Another interesting problem (of a dynamic particle coupled to the 1D
metal) arises, for instance, if one studies the core--hole recoil.
A very recent investigation of this problem was performed in the paper
\cite{caldeira}. The neglect of the backscattering, however,
done in this paper, seems to be crucial since it is known (see, e.g.,
Ref.\cite{forward}) that the forward scattering alone in not
sufficient to provide the orthogonality of metal wave--functions
adjusted to different positions of the particle \cite{notepolaron}.
It is the backscattering, which provides the orthogonality and leads
to a dissipative dynamics of the particle, and which is, therefore,
expected to be the most important process (giving the leading
contribution, for instance, to the temperature dependence of the
diffusion coefficient). In any case, it is difficult to imagine that
the particle is localized. The core--hole recoil should always
broaden Fermi edge singularities. Note also that in real experiments most of
the core--holes, being created in the bulk semiconductor, are not
affected by the confining potential and therefore have more recoil
possibilities (i.e. the core--hole can escape in any direction).

For the case of spinning electrons an analysis, very similar to that one
presented above, has been carried out \cite{K&Fbis,F&N,GL,AFFL}.
There are now two exponents, $K_\rho$ and $K_\sigma$, governing the
asymptotics of spin--spin and density--density correlation functions
\cite{Solyom&Emery}. The renormalization group study of the
impurity potential leads to qualitative conclusions, similar
to those of the spinless case: the impurity potential is predicted
to be renormalized by the interaction either to zero or to infinity,
depending on the values of the parameters $K_\rho$ and $K_\sigma$.
This implies, for instance, that the zero--temperature conductance
is either perfect or vanishing, except the only one point
(corresponding to the non--interacting electrons) where it
depends on the impurity potential.
As to magnetic field effects (see also Ref.\cite{GL}),
one clearly should expect a crossover
behavior of the conductance temperature dependence with the crossover
temperature of the order of the Zeeman splitting, $2\mu_B H$.
For the case of a short--range electron--electron interaction
(effectively screened Coulomb repulsion),
one might expect a large positive magnetoresistance.
On the contrary, for
the case of a long--range interaction, the conductance would be less
sensitive to the magnetic field. Indeed, the short--range part of the
electron--electron
interaction suppresses the conductance in
zero magnetic field but it is effectively absent in the case of  strong
(polarizing) magnetic field (since, in the latter case,
the problem essentially reduces to the problem of spinless fermions);
whereas the long--range part of the interaction effectively
suppresses conductance for any
magnetic field.

In many experimental set--ups, not just the lowest 1D band but several bands
are filled. However, except the case of two bands, which is
equivalent to the intensively studied problem of two chains\cite{michele},
there are not many theoretical results on the many--band case available.
To my knowledge, it is only the so--called multi--component Luttinger
liquid fixed point which was investigated
(see Ref.\cite{multy} for the pure case;
some impurity effects at this fixed point are discussed
in the paper \cite{our}). It is quite
clear, however, that for realistic systems this fixed point is
unstable. One can nevertheless hypothesize what the ground--state (of the pure
system) looks like by making use of the scale invariance ideas. Indeed,
the model containing bands with different Fermi velocities coupled by
the interaction is not scale invariant. Therefore, one may expect that
there are two things which can happen
under scaling transformations: either the bands become decoupled while
maintaining different Fermi velocities or the bands remain coupled
but the difference in velocities shrinks to zero and the model becomes
symmetric. Thus, the whole many--band system will decouple into
symmetric subsets of bands. This scenario
is true for the simplest case of two bands \cite{michele},
and it seems to be
confirmed by preliminary renormalization group calculations
for the many--band case \cite{multybis} but further studies are required.

\appendix
\section*{}
\label{app-A}

In this appendix I give the derivation of
the Eq.(\ref{formula}). The single electron Schr\"{o}dinger equation,
\be
\left\{ \varepsilon_{0} \left( -i\partial_{x}
\right) +V(x)
-\varepsilon \right\} \psi \left[ x \right] =0\; ,
\label{appschro}
\ee
where
$\varepsilon_{0}(p)$ is the dispersion law,
should be solved under the twisted boundary condition:
\be
\psi (x+L)=e^{i\varphi}\psi (x)\; .
\label{apptwiced}
\ee
It was recognized already in the pioneering papers \cite{early}
that this problem is equivalent to the Bloch functions problem, where
the whole ring plays the role of the elementary cell
and $\varphi$ - the role of the quasimomentum.
The ground--state energy as a function of the flux
is given by the sum
\be
E_{0}(\varphi )=\sum_{\lambda}\varepsilon_{\lambda}(\varphi )
\label{appen}
\ee
over $N$ lowest values of the band spectrum at fixed $\varphi$.

Suppose, for clarity, that the potential $V(x)$ is localized
within the region of a radius $a<L$
(actually, the relation between $a$ and $L$ can be shown
to be unimportant).
The wave-function $\psi (x)$ can be then written in the form
$\psi (x)=Ae^{ipx}+Be^{-ipx}$ from the left of the potential
and in the form $\psi (x)=Ce^{ipx}+De^{-ipx}$ from the right of it.
The coefficients $C$ and $D$ can be expressed through $A$ and $B$
by making use of the scattering data of $V(x)$ (transfer matrix).
The twisted boundary condition  gives then
(after elementary algebra) the equations for the spectrum:
\be
pL =  \Phi_{+}(p,\varphi ) \;\;\;( n=0);\;\;\;\;\;
pL = 2\pi n + \Phi_{\pm}(p,\varphi )\;\;\; ( n=1,2,...);  \\
\label{appspectrum}
\ee
where $\Phi_{\pm}(p,\varphi )=\delta_f (p) \pm F(p,\varphi )$ and
the function $F$ is defined by
\be
F(p,\varphi )=
\cos^{-1} \left( \sqrt{D(p)}\cos \varphi \right)
\label{appF}
\ee
and $\delta_f (p)$ is the same forward scattering phase
as the one in the Eq.(\ref{expabis}).

To make progress with the Eq.(\ref{appspectrum}), the idea is to expand the
solution in $1/L$.
This is simplest when $a<<L$ (e.g. $V(x)$ represents a single
scatterer) and $\varepsilon_{0}(p)=p^{2}/2m^{*}$. Then one can write:
\be
p_{n}=k_{n}+\frac{1}{L} \Phi_{\pm}(k_{n})+
\frac{1}{L^{2}} \Phi_{\pm}(k_{n})
\frac{\partial\Phi_{\pm}(k_{n})}{\partial k}
+O\left( \frac{1}{L^{3}}\right) ;\;\;\; \; k_{n}=2\pi n/L
\label{appp}
\ee
For the ground--state energy we therefore have:
\be
\left.
E_{0}=\frac{1}{m^{*}}\sum_{n}\left\{ k^{2}+\frac{2}{L}k\delta (k)+
\frac{1}{2L^{2}}\frac{\partial}{\partial k} \left[ k
\sum_{\pm}\Phi_{\pm}^{2}(\varphi ,k)\right]
+O\left(\frac{1}{L^{3}}\right) \right\} \right| _{k=k_{n}}
\label{appenbis}
\ee
Here the first term is the ground--state energy in the absence of
the potential and flux (it is proportional to the volume: $\sim L$).
The second term ($\sim 1$) is the energy difference due the scattering
potential (in agreement with the Fumi theorem\cite{FT}), which is
flux independent. Thus, the effect of the flux is of the order of
$1/L$ and is given by:
\bea
\Delta E_{0}(\varphi )&=&E_{0}(\varphi )-E_{0}(0)=
\left. \sum_{n}\left\{
\frac{1}{2L^{2}}\frac{\partial}{\partial k} \left[ k
\sum_{\pm}\Phi_{\pm}^{2}(\varphi ,k)-
\Phi_{\pm}^{2}(0,k)\right]
+O\left(\frac{1}{L^{3}}\right) \right\}
\right| _{k=k_{n}} = \nonumber \\
&=&\frac{v_{_F}}{2\pi L} \left\{
F^{2}(p_{F},\varphi )-F^{2}(p_{F},0)\right\}
+O\left(\frac{1}{L^{2}}\right)\;.
\label{appdeltaen}
\eea
For $N=even$ one additional particle on top of the spectrum
contributes to Eq.(\ref{appenbis}) the term:
\be
\Delta E_{0}^{par}(\varphi )=-{v_{F} \over L} \left\{
F(p_{F},\varphi )-F(p_{F},0) \right\}
\label{appparity}
\ee
In fact, each particle contributes to the flux dependence of the
energy a term $\sim 1/L$, but the contributions of the particles with
quantum numbers $(n,+)$ and $(n,-)$, which would correspond to the
momenta $+p$ and $-p$ for $V=0$, almost cancel each other,
and the entire contribution of the Fermi sea is again of the
order of $1/L$ and converges actually just at the Fermi surface
(Eq.(\ref{appdeltaen})).
For even $N$, the particle on the top, i.e.
in the state $(N/2,-)$, does not have a partner in the state
$(N/2,+)$, so the contribution of this single particle is $\sim1/L$;
that gives rise to the parity effect, Eq.(\ref{appparity}).
The "Landauer type" formula for the persistent current, Eq.(\ref{formula}),
follows by deriving Eqs(\ref{appdeltaen},\ref{appparity}) with
respect to the flux\cite{parity}.

\newpage
\begin{center}
{\bf Figure Captions}
\end{center}
{\bf Fig.1}. The electron band structure in confining potential.\\
{\bf Fig.2}. Energy dependence of the transmission coefficient
renormalized by the interaction.\\
{\bf Fig.3}. Qualitative behavior of the $\beta$-function,
Eq.\protect\ref{beta}\protect .\\
{\bf Fig.4}. Impurity potential scaling flow phase diagram.\\
{\bf Fig.5}. Qualitative temperature dependence of the
persistent current amplitude
for the interacting electron gas with an impurity.


\begin{references}
\bibitem{backV}
Strictly speaking, $V$ is understood here to be the backscattering part
(the $2p_F$ Fourier component) of the impurity potential $V(x)$.
\bibitem{notexp}
Note that the exponent $K$ is not renormalized, $dK/d\xi =0$, since $K$ is a
bulk property which can not be changed by a single impurity.
\bibitem{notetime}
i.e. for $t>>t_0$, where the crossover time is estimated as
\protect\cite{mykol}\protect: $t_{0}^{-1} \sim W(V/W)^{|2/K-2|}$.
\end{references}
\end{document}